\documentclass[twocolumn]{aa}

\usepackage{epsfig,graphicx,natbib}
\usepackage{longtable,amssymb}

\def\Tcmb{\hbox{$T_\mathrm{CMB}$}}
\def\Trot{\hbox{$T_\mathrm{rot}$}}
\def\Tkin{\hbox{$T_\mathrm{kin}$}}
\def\Ncol{\hbox{$N_\mathrm{col}$}}
\def\NLTE{\hbox{$N_\mathrm{LTE}$}}
\def\nH2{\hbox{$n_\mathrm{H_2}$}}
\def\V0{\hbox{$\mathrm{V}_0$}}
\def\DV{\hbox{$\Delta \mathrm{V}$}}

\begin{document}

\title{A precise and accurate determination of the cosmic microwave background temperature at $z$=0.89}

\author{S. Muller \inst{1}
\and A. Beelen \inst{2}
\and J. H. Black \inst{1}
\and S. J. Curran \inst{3,4}
\and C. Horellou \inst{1}
\and S. Aalto \inst{1}
\and F. Combes \inst{5}\and M. Gu\'elin \inst{6,7}
\and C. Henkel \inst{8,9}
}

\institute{
Department of Earth and Space Sciences, Chalmers University of Technology, Onsala Space Observatory, SE 439 92 Onsala, Sweden
\and Institut d'Astrophysique Spatiale, B\^at. 121, Universit\'e Paris-Sud, 91405 Orsay Cedex, France
\and Sydney Institute for Astronomy, School of Physics, The University of Sydney, NSW 2006, Australia
\and ARC Centre of Excellence for All-sky Astrophysics (CAASTRO)
\and Observatoire de Paris, LERMA, CNRS, 61 Av. de l'Observatoire, 75014 Paris, France
\and Institut de Radioastronomie Millim\'etrique, 300, rue de la piscine, 38406 St Martin d'H\`eres, France 
\and Ecole Normale Sup\'erieure/LERMA, 24 rue Lhomond, 75005 Paris, France
\and Max-Planck-Institut f\"ur Radioastonomie, Auf dem H\"ugel 69, D-53121 Bonn, Germany
\and Astron. Dept., King Abdulaziz University, P.O. Box 80203, Jeddah, Saudi Arabia
}

\date {Received / Accepted}

\titlerunning{$T_{\rm CMB}$ at $z$=0.89}
\authorrunning{Muller et al. 2013}

\abstract{According to the Big Bang theory and as a consequence of adiabatic expansion of the Universe,
the temperature of the cosmic microwave background (CMB) increases
linearly with redshift. This relation is, however, poorly explored, and
detection of any deviation would directly lead to (astro-)physics beyond the standard model.}
{We aim at measuring the temperature of the CMB with an accuracy of a few percent
at $z$=0.89 toward the molecular absorber in the galaxy lensing the quasar PKS\,1830$-$211.}
{We adopt a Monte-Carlo Markov Chain approach, coupled with predictions from the non-LTE radiative transfer code RADEX,
to solve the excitation of a set of various molecular species directly from their spectra.}
{We determine \Tcmb=5.08$\pm$0.10~K at 68\% confidence level. Our measurement is consistent with the value \Tcmb=5.14~K
predicted by the standard cosmological model with adiabatic expansion of the Universe.
This is the most precise determination of \Tcmb\ at $z$$>$0 to date.}
{}
\keywords{cosmology: observations - cosmic background radiation - cosmological parameters - quasars: absorption lines
- quasars: individual: PKS\,1830$-$211}

\maketitle

\section{Introduction}

\subsection{The CMB temperature as a function of redshift}

The cosmic microwave background (CMB) is one of the pillars of the Big Bang theory.
This radiation field is a relic of the epoch when matter and photons decoupled at $z$$\sim$1100 and
fills up the entire sky with a photon density of $\sim$400 cm$^{-3}$.
Since decoupling, the CMB photons have cooled
down with the expansion of the Universe. Observations with the COBE satellite have demonstrated that
the CMB corresponds to a nearly perfect black body, characterized by a temperature $T_0$ at $z$=0,
which is measured with very high accuracy, $T_0$=$2.72548 \pm 0.00057$~K (\citealt{fix09}).

Measuring the CMB temperature at high redshift has considerable cosmological interests, in {\em 1)}
demonstrating that the CMB radiation is universal (Equivalence principle) and {\em 2)} tracing the
evolution of its temperature with redshift, \Tcmb($z$).
Adiabatic expansion predicts that the CMB temperature evolution is proportional to (1+$z$).
Alternative cosmologies, such as decaying dark energy models ({\em e.g.} \citealt{lim96,lim00,jet11})
where dark energy can interact with matter via creation of photons, affecting the CMB spectrum,
predict a deviation from this simple law with $T$($z$)=$T_0$$\times$(1+$z$)$^{(1-\alpha)}$.
Precise measurements of \Tcmb\ at high redshift can therefore constrain those theories.

Two methods can be used to probe the CMB temperature at $z$$>$0.
The first one is based on multi-frequency Sunyaev-Zeldovich (S-Z)
observations toward galaxy clusters (see {\em e.g.} \citealt{hor05}).
Inverse Compton scattering by free electrons in a cluster of galaxies induces a distortion of the CMB spectrum,
which can be used to constrain the CMB temperature at the location of the cluster.
\cite{bat02} reported the first determination of \Tcmb($z$)
toward the Coma ($z$=0.02) and Abell~2163 ($z$=0.2) clusters.
More recently, this method was applied to a sample of 13 clusters between $z=$0.023 and $z$=0.55
by \cite{luz09}, who found $\alpha$=0.024$_{-0.024}^{+0.068}$, thus a \Tcmb--$z$ law in agreement with
the standard model. One limitation of the S-Z method is the scarcity of clusters at high-redshift ($z$$>$1).
\cite{mar12} discuss prospects in the near future based on Planck data. 

The second method relies on spectroscopic studies of lines in absorption against quasars and their
excitation analysis. Most of the measurements at high redshift have used UV spectroscopy of atomic
species (\ion{C}{i}: \citealt{mey86,son94b,ge97,rot99}; \ion{C}{ii}: \citealt{son94a,lu96,mol02};
\ion{C}{i} \& \ion{C}{ii}: \citealt{sri00}). Strictly speaking, these measurements are all upper
limits on \Tcmb, as the contributions from other local sources of excitation (collisions, local
radiation field) are yet largely uncertain and poorly constrained, and have to be accounted for.
More recently, \cite{not11} observed electronic transitions of CO absorption toward several quasars
at redshift between 1.7 and 2.7. Together with the previous measurements of \Tcmb($z$) in the
literature, they could obtain the tightest constraint to date on the temperature evolution of the
CMB, $\alpha$=$-$0.007$\pm$0.027, consistent with adiabatic expansion of the Universe. Nevertheless,
all the studies so far have relied on a limited number of transitions/species, introducing a possible
bias in the determination of excitation conditions. Also, the UV spectroscopy technique ({\em e.g.}
of \ion{C}{i} and \ion{C}{ii}) is not suitable below z$\sim$1, due to the atmospheric cut-off around 300~nm.

The use of radio-mm molecular absorbers to determine \Tcmb\ at $z$$>$0 is another particularly
attractive method, due to the tighter constraints on physical conditions and local excitation when a
variety of molecular species is detected. In this paper, we follow this approach, with the aim to
determine the CMB temperature from multi-transition multi-species excitation analysis in the $z$=0.89
radio-mm molecular-rich absorber toward the quasar PKS\,1830$-$211.

\subsection{The $z$=0.89 molecular-rich absorber toward PKS\,1830$-$211} \label{sect-pks}

PKS\,1830$-$211 is a radio-loud blazar at a redshift $z$=2.5 (\citealt{lid99}), gravitationally
lensed by a $z$=0.88582 almost face-on spiral galaxy (\citealt{wik96,win02}). The presence of an
additional galaxy at $z$=0.19 on the line of sight was inferred from \ion{H}{i} absorption (\citealt{lov96}).
The lensed image of the quasar appears in the radio continuum as a textbook example of a complete
Einstein ring of 1$''$ in diameter, plus two bright and compact components, located to the NE and
SW of the ring (\citealt{jau91}). The Einstein ring has a steep spectral index and grows fainter
with frequency, until invisible in the millimeter. The NE and SW images of the quasar are marginally resolved in 43~GHz VLBI
observations with 0.5~mas beam resolution by \cite{jin03}. At the distance of the $z$=0.89
galaxy, their angular size corresponds to a scale of $\sim$1~pc at mm wavelengths. Most interestingly,
the line of sight toward both images intercepts molecular clouds located in the spiral arms of the intervening $z$=0.89
galaxy, which results in remarkable molecular absorptions (\citealt{wik98,men99,mul06}). In fact,
PKS\,1830$-$211 is the strongest of only four redshifted mm-molecular absorbers known to date (\citealt{com08}),
the one at the highest redshift, and the most molecule-rich (\citealt{mul11}). As such,
PKS\,1830$-$211 is a target of choice for using molecules as cosmological probes, for example of
the CMB temperature, of the constancy of fundamental constants, or of the chemical evolution of the
Universe (see {\em e.g.} \citealt{hen09,mul11b}).

The activity of the blazar leads to time variations of its flux, with a time delay between the NE
and SW images of about 25~days (\citealt{lov98,wik99}), as well as changes in the morphology and
barycenter of the continuum emission, seen in VLBI observations (\citealt{gar97,jin03}). \cite{nai05}
interpret the changes in the apparent distance between the NE and SW images by the recurrent ejection
of plasmons along a helical jet in the blazar. Consequently, and of great importance for our study,
the continuum emission of the two images is drifting behind the absorbing gas, at the rhythm of plasmon
ejections. \cite{jin03} measured drifts up to 200~$\mu$as (1.6~pc in the lens plane at $z$=0.89) over
eight months. The changes of sightline result in a time variation of the molecular absorption profiles,
as revealed by the long-term monitoring of the HCO$^+$/HCN 1-0 lines by \cite{mul08}. 
This time variation could hamper studies which rely on the comparison of different molecular transitions,
such as constraints on the constancy of fundamental constants, multi-transition excitation analysis, 
as well as abundance and ratio measurements.

\section{Observations and data reduction} \label{Obs}

Our main goal was to observe several rotational transitions of various molecular species in order to derive
their excitation and constrain the CMB temperature toward PKS\,1830$-$211. In particular, to take advantage
of the two SW/NE lines of sight through the intervening galaxy, which are independent (on either side of the
bulge, at galactocentric distances of $\sim$2~kpc and $\sim$4~kpc, respectively) and likely to have different
excitation conditions, giving a robust measurement of \Tcmb\ at $z$=0.89.

Toward the NE line of sight, only a handful of species have been detected (\citealt{mul11}): we
have chosen to focus on the HCO$^+$, HCN, and HNC J=1-0, 2-1, and 3-2 transitions, which are all
optically thin. These same transitions
are saturated or optically thick toward the SW line of sight, but their isotopic variants, observable
in the same tunings, can be used instead. Several other species, such as HOC$^+$ and SiO, also have
rotational transitions in the same observable bands.

In order to cover the different rotational transitions of our targeted molecular species, we used
the Australia Telescope Compact Array (ATCA) at 7~mm and 3~mm and the Plateau de Bure interferometer
(PdBI) at 3~mm and 2~mm. Details on the respective observations and specific data calibration are
given hereafter in \S.\ref{atca} and \S.\ref{pdbi}, and a summary of the observations is given in
Table~\ref{tab-journal}. Unfortunately, because of scheduling and weather constraints the ATCA and
PdBI observations could not be carried out simultaneously, which raises the concern of time variations
of the source between the observations. This problem is discussed in \S.\ref{Ibg}.

The full final spectra are shown in Fig.\ref{fig-spec-full} and \ref{fig-spec-zoom}. A crude comparison
of the new data and ATCA spectra obtained in 2009/2010 shows that significant variations have occurred
(Fig.\ref{fig-spec-zoom}): while the intensities of the SW absorption lines ($v$=0~km\,s$^{-1}$) have
increased by a factor of $\sim$2 for most species, the absorption depths of the HCO$^+$ and HCN lines
toward the NE image ($v$=$-$147~km\,s$^{-1}$ \footnote{All velocities throughout this work refer
to the heliocentric reference frame, adopting a redshift $z$=0.88582}) have significantly decreased.
The resulting low signal-to-noise ratio on these lines and the continuum baseline uncertainty due to
the wings of the corresponding saturated lines from the SW component (see Fig.\ref{fig-spec-zoom})
prevent us from obtaining a robust fit, which hampers our initial goal to derive \Tcmb\ in several
independent lines of sight. 

Other velocity components ({\em e.g.} at velocity $-$224~km\,s$^{-1}$ and +170~km\,s$^{-1}$,
see \citealt{mul11}), are also too weak for a meaningful analysis.

\begin{table*}[ht] 
\caption{Summary of the observations.} \label{tab-journal}
\begin{center} \begin{tabular}{ccccccccc}
\hline
Date & Telescope & Band & Frequency & $\delta$V $^a$ & $\sigma$ $^b$ & $I_{bg}$(SW) $^c$ & R $^d$ & Notes \\
     &           &      & (GHz)     & (km\,s$^{-1}$)          & (10$^{-3}$)    &         &    &  \\
\hline
2011 Jul. 24 & PdBI & 3~mm & 93.0 -- 96.6   & --/2.0 $^e$ & --/2.3 $^e$ & 0.416 $\pm$ 0.006 & 1.40 $\pm$ 0.03 & Total flux=2.6~Jy \\
2011 Jul. 27 & ATCA & 7~mm & 44.8 -- 48.9   & 6.3 & 2.0         & 0.402 $\pm$ 0.003 & 1.49 $\pm$ 0.01 & \\
2011 Jul. 27 & ATCA & 3~mm & 90.8 -- 96.4   & 3.2 & 2.8         & 0.397 $\pm$ 0.004 & 1.52 $\pm$ 0.02 & \\
2011 Aug. 17 & PdBI & 2~mm & 140.2 -- 143.8 & 4.1/1.3 $^e$ & 4.5/6.8 $^e$ & 0.437 $\pm$ 0.010 & 1.29 $\pm$ 0.05  & Total flux=2.1~Jy \\
\hline
\end{tabular} 
\tablefoot{$^a$ Velocity resolution for the HCO$^+$ lines.
$^b$ Noise level normalized to the total continuum (NE+SW images).
$^c$ Fraction of the total continuum intensity corresponding to the SW image,
measured on the saturated part of the spectra between $-7$~km\,s$^{-1}$ and +7~km\,s$^{-1}$.
$^d$ Magnification factor R=$I_{bg}$(NE)/$I_{bg}$(SW), assuming no flux comes from other components.
$^e$ Corresponding to the WIDEX and narrow-band correlator output, respectively.
}
\end{center} \end{table*}

\subsection{Australia Telescope Compact Array (ATCA)} \label{atca}

Observations were carried out with the ATCA on 2011 July 27, with four antennas in a compact configuration.
The longest baseline was $\sim$150~m, providing an angular resolution $>$9$''$ at 7~mm and $>$4$''$ at 3~mm.
The continuum emission (1$''$ Einstein ring and NE/SW images) of PKS\,1830$-$211 was therefore not resolved.

The CABB correlator was set up to provide a bandwidth of 2~GHz with a spectral channel spacing of 1~MHz in each
sideband. We used two partly overlapping tunings at 7~mm and three at 3~mm to cover the frequency ranges given
in Table~\ref{tab-journal}. The on-source integration time was 0.5~h/tuning at 7~mm and 1.5~h/tuning at 3~mm.
The total allocated time, including calibration and overheads, was 8~h. The receivers were tuned in dual
polarisation mode. Both polarisations were averaged after the data calibration. The bandpass calibration was
performed on the bright quasar 1921$-$293 for each tuning.

The data reduction was done using MIRIAD~\footnote{http://www.atnf.csiro.au/computing/software/miriad/}
(\citealt{sau95}). Source visibilities were self-calibrated and the (unresolved) continuum level normalized to
unity. Spectra were then extracted from the calibrated visibilities following the same method as described by
\cite{mul11}. A fixed Doppler track correction of +13~km\,s$^{-1}$ was applied to all data. The drift in
velocity being less than 0.1~km\,s$^{-1}$/hour, the effect introduced by adopting a fixed correction is
negligible considering the spectral resolution.


In complement to this dataset, we use previous ATCA 7~mm observations from 2009 September 1 and 2
published in \cite{mul11}, specifically for transitions of c-C$_3$H$_2$, CH$_3$CN, SO, and HC$_3$N
(see \S.\ref{sec:RADEXmodel}).
These observations were obtained and reduced following the same procedure as described above.

\subsection{Plateau de Bure interferometer (PdBI)} \label{pdbi}

Observations were obtained with the PdBI on 2011 July 24 at 3~mm, and on 2011 August 17 at 2~mm,
with five antennas in a compact array. All baselines were shorter than 100~m, so that the continuum
emission of PKS\,1830$-$211 was not resolved, as for ATCA observations.

\begin{figure*}[h] \begin{center}
\includegraphics[width=\textwidth]{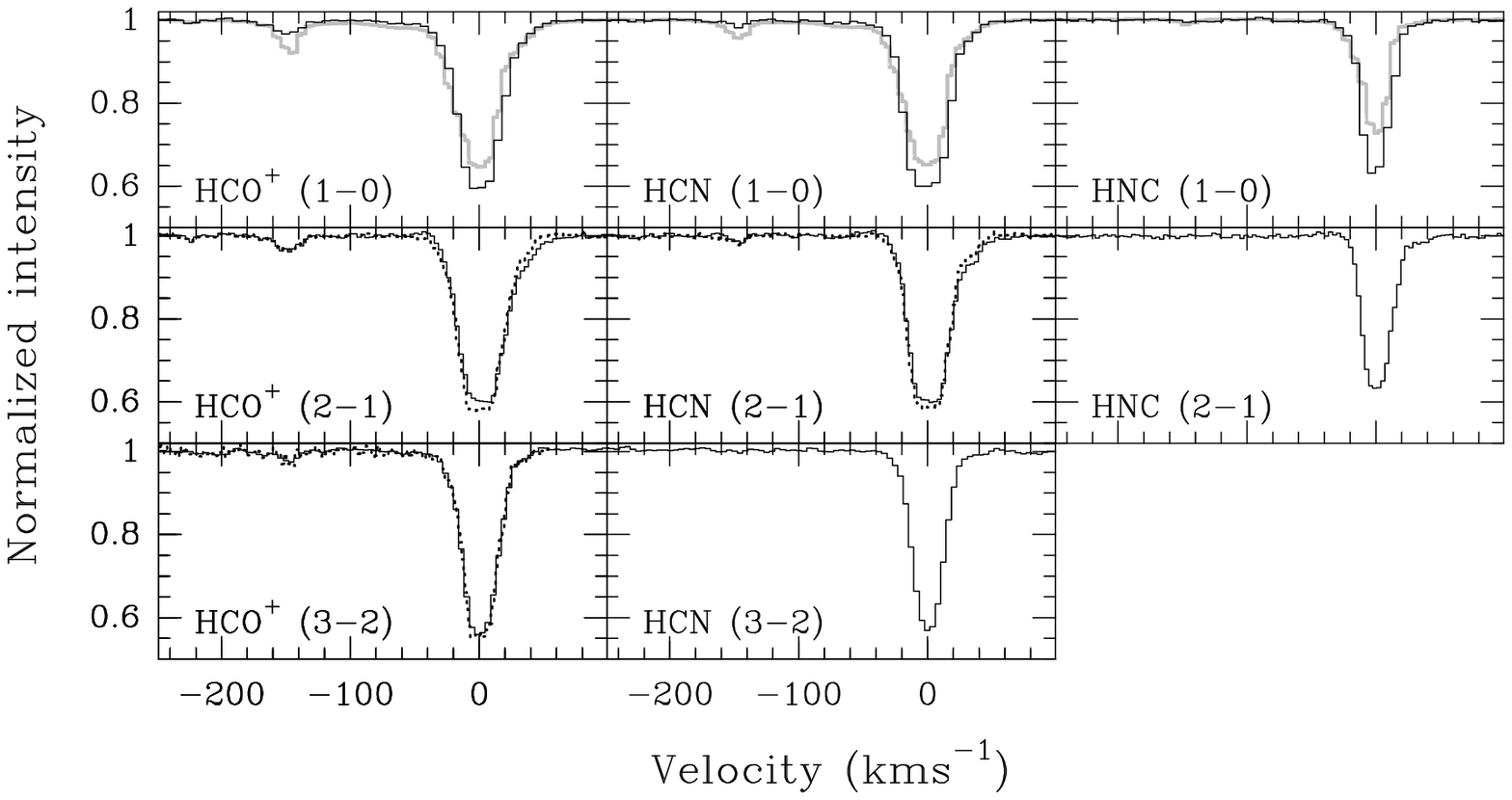}
\caption{Spectra of the strong absorption of HCO$^+$, HCN and HNC transitions observed in 2011
with ATCA (full black line).
The dotted spectra correspond to PdBI observations with the narrow-band correlator,
{\em i.e.,} at better velocity resolution (see Table~\ref{tab-journal}).
The J=1-0 spectra observed in our previous ATCA 7~mm survey (\citealt{mul11}) are shown in light grey.
The intensity is normalized to the total continuum flux (NE plus SW images).}
\label{fig-spec-strong}
\end{center} \end{figure*}

The wide-band correlator WIDEX was used, and provided a 3.6~GHz bandwidth with a spectral resolution
of 1.95~MHz. In addition, the narrow-band correlator was configured to select two 160~MHz spectral
bands with a better spectral resolution of 0.625~MHz, with dual polarisation. At 3~mm, these narrow
bands were set to cover the 0~km\,s$^{-1}$ and $-$147~km\,s$^{-1}$ absorption lines of both HCO$^+$
and HCN 2-1. At 2~mm, the narrow-band IF range allowed us to cover only the 0~km\,s$^{-1}$ and
$-$147~km\,s$^{-1}$ components of the HCO$^+$ 3-2 line and the $-$147~km\,s$^{-1}$ component of the
HCN 3-2 line (see the spectra in Fig.\ref{fig-spec-strong}).

The bandpass was calibrated on the bright quasars 1749+096 (at 3~mm) and 3C454.3 (at 2~mm). We followed
the standard calibration procedure in CLIC~\footnote{http://www.iram.fr/IRAMFR/GILDAS/},
using the ``Self-Calibration on point source'' routine.
The spectra were then exported to CLASS~$^2$, where we divided the spectral baseline with polynomials of degree
two for the narrow-band spectra, and degree 12 or 24 depending on the case for the full 3.6~GHz WIDEX bands,
to correct for residual ripples in the spectral baseline. The target absorption lines are too narrow
to be seriously affected by these polynomial orders.

Features corresponding to O$_3$ atmospheric lines were identified in the final spectra.
In particular, we flagged a region of 200~MHz around the 142.175~GHz O$_3$ line in the 2~mm data.


\subsection{Continuum illumination} \label{Ibg}

The absorption intensities ($I_\nu$), measured from the total continuum level, are converted to optical
depths ($\tau_\nu$) according to:
\begin{equation}
 \tau_\nu = - \ln{ \left ( 1-\frac{I_\nu}{f_c I_{bg}} \right ), }
\label{eq:tau}
\end{equation}
\noindent where $I_{bg}$  and $f_c$ are the fraction of the total continuum intensity (normalized to unity) and
the continuum source covering factor, corresponding to each individual image of the quasar, respectively.
We adopt a value $f_c$=1 (see \citealt{wik98,mul08}).

At radio wavelengths, the continuum emission of the background quasar is dominated by two bright and
compact images to the NE and to the SW of the 1$''$~Einstein ring (see \S\ref{sect-pks}). These NE and
SW images are not resolved in our observations, so that we cannot measure directly their respective
fluxes $I_{bg}$(NE) and $I_{bg}$(SW). However, the HCO$^+$ and HCN lines appear to be flat-bottomed near
$v$=0~km\,s$^{-1}$, and both at the same intensity for a given epoch (Fig.\ref{fig-spec-strong}), which
strongly suggests that they are saturated. The detection of the $^{13}$C variants of these species (and
even of other less abundant isotopologues) clearly implies that they are optically thick, with opacities
higher than $\tau$=10. Because we know that the absorption near $v$=0~km\,s$^{-1}$ occurs in front of
the SW image (\citealt{wik98,mul06}), we can measure $I_{bg}$(SW) from the saturation level of the
HCO$^+$ and HCN lines between $v$=$-$7~km\,s$^{-1}$ and $v$=+7~km\,s$^{-1}$, under the assumption $f_c$=1.
The results are given in Table~\ref{tab-journal}. The comparison of the HCO$^+$~J=3-2 spectra obtained
with the WIDEX correlator (velocity resolution $\delta V$ of $\sim$4~km\,s$^{-1}$) and with the
narrow-band correlator ($\delta V$$\sim$1~km\,s$^{-1}$) shows that the limited resolution in velocity
does not affect significantly the determination of the saturation level.

The measured values of $I_{bg}$(SW) differ between the three epochs, with the biggest change on the last
measurement with the PdBI, obtained 20~days after the ATCA observations. As the saturation level does not
depend on the absolute flux density calibration (the intensities are normalized to the total continuum level),
we believe that these changes are significant, and not of instrumental origin. 
They could be caused either by a change of the NE/SW ratio due to the time delay of $\sim$25~days between
the two images of the quasar (\citealt{lov98, wik99}), or to a change of the source covering factor if the
sightline moved between the observations. Changes of the continuum morphology, related to the jet activity of
the blazar, are clearly seen within a time span of two weeks (\citealt{jin03}, see also \citealt{gar97}), but
monitoring on a shorter timescale is still lacking, to the best of our knowledge.

The changes seen between July and August 2011
are unfortunate for our purpose, because the lines of sight through the intervening galaxy
-- hence the column density and physical conditions of the absorbing gas -- might have changed. This
emphasizes the need to conduct the observations within a short time interval
when the goal is to combine different observations toward PKS\,1830$-$211.
In the rest of the paper, we will treat only the ATCA observations of the lines corresponding to the SW component,
and assume that no significant intra-day variations occur.

In order to take into account the uncertainty on the continuum illumination $\Delta I_{bg}$, we have added
quadratically to the fit uncertainties of the integrated opacities a term calculated as:
\begin{equation}
\Delta \left ( \int \tau dv \right ) = \left ( \int \tau dv \right ) \times \frac{\Delta I_{bg}}{I_{bg}} \times F(I/I_{bg}),
\label{eq:dIbg}
\end{equation}
\noindent where the function $F(x) = x / ((x-1)\ln{(1-x)})$, as obtained from the differentiation of
Eq.\ref{eq:tau} with respect to $I_{bg}$. This correction is simply equivalent to
$\frac{\Delta \tau}{\tau}$=$\frac{\Delta I_{bg}}{I_{bg}}$ for optically thin lines.

The magnification ratios R=NE/SW, which we take as the fraction (1$-$$I_{bg}$(SW))/$I_{bg}$(SW), are listed
in Table~\ref{tab-journal}. They are within the range of values compiled by \cite{mul08}, using the same
measurement method of the saturation level of the HCO$^+$ SW absorption.

\section{Analysis}

In this section we describe the analysis of the spectral data, starting from a simple approach
to derive rotation temperatures. We then investigate the effects of local physical conditions on the
line excitation. Finally, we provide a global solution of the problem using a Monte-Carlo Markov Chain
method, from which we derive our final measurement of \Tcmb\ at $z$=0.89.

\subsection{Rotation temperatures} \label{TROT}

First, we focus on deriving the rotation temperature of the observed species
assuming that the excitation is dominated by radiative coupling with the CMB (low-density case).
Under this condition, the population of the energy levels follows a Boltzmann distribution,
for which the rotation temperature equals \Tcmb.
The rotation temperature \Trot\ between two energy levels (upper $u$ and lower $l$)
is defined by the Boltzmann equation:
\begin{equation}
\frac{n_u}{n_l} = \frac{g_u}{g_l} \exp{ \left ( - \frac{h \nu_{ul}}{k \Trot} \right ) }
\label{eq:trot}
\end{equation}
\noindent where $n_u$ and $n_l$ are the populations, $g_u$ and $g_l$ the degeneracies,
of the upper and lower level, respectively, connected by a transition at frequency $\nu_{ul}$.
The column density \NLTE~\footnote{We refer to this estimate as the LTE column density, although
strictly speaking, we simply assume that the excitation is dominated by the CMB photons.}
can then be derived as:
\begin{equation}
\NLTE = \frac{3h}{8\pi^3\mu^2S_{ul}}\frac{Q(\Trot)\exp{\left ( \frac{E_l}{k_B \Trot} \right )}}{\left [ 1-\exp{\left ( - \frac{h\nu}{k_{B} \Trot} \right )} \right ]} \int \tau dv, 
\label{eq:ncol}
\end{equation}
\noindent where $Q(\Trot)$ is the partition function, $E_{l}$ the energy of the lower level
with respect to ground state, $\mu$ the dipole moment, $\nu$ is the frequency, $S_{ul}$ the
line strength and $\int \tau dv$ the integrated opacity of the line.

The kinetic temperature of the gas along the SW line of sight is of the order of $\sim$80~K,
as indicated by observations of the symmetric top molecules NH$_3$ and CH$_3$CN
(\citealt{hen08, mul11}). Because the density is moderate (of the order of $10^3$~cm$^{-3}$,
\citealt{hen09}), the gas is sub-thermally excited and the excitation mostly coupled radiatively
with the CMB. Hence, the rotation temperatures are expected to be close to (only slightly higher
than, see \S\ref{sect-effectsoflocalconditions}) \Tcmb. The rotation temperatures of several
species have been derived by \cite{com99}, \cite{men99}, \cite{hen09}, and \cite{mul11} in the
SW line of sight, and are consistent with this picture.

\begin{figure*}[ht] \begin{center}
\includegraphics[width=\textwidth]{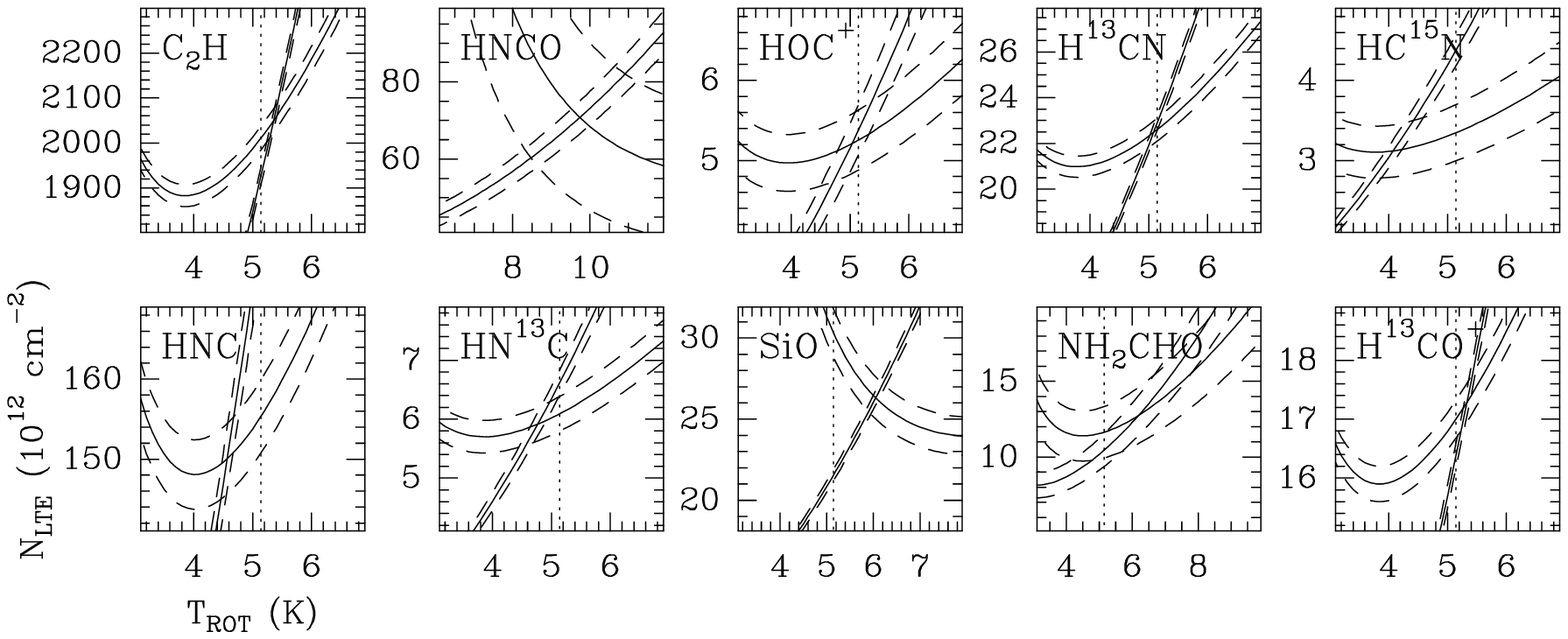}
\caption{Evolution of the column density (\NLTE) as a function of the rotation temperature (\Trot)
for the transitions of different molecular species observed in 2011 (Table~\ref{tab-lines}), according
to Eq.\ref{eq:ncol}. Each curve corresponds to a different transition and the dashed curves delimit
the 1$\sigma$ uncertainties. The curve intersections give the values of \Ncol\ and \Trot. 
The vertical dotted line indicates \Tcmb=5.14~K.}
\label{fig-trot}
\end{center} \end{figure*}

We note that the Rayleigh-Jeans approximation does not hold for $\nu$$>$100~GHz transitions
when the temperature is low ($\sim$5~K), and the rotation diagram method (as done in \citealt{mul11}
for lower frequency data) is not applicable to derive the rotation temperature here. Instead, we
use Eq.\ref{eq:ncol}, and plot the \NLTE--\Trot\ curves for the different observed transitions
(Fig.\ref{fig-trot}). These curves have slightly different sensitivities to the temperature. Their
intersection gives the rotation temperature and column density. The results are given in
Table~\ref{tab-trot}. As expected, the rotation temperatures are all close to $\sim$5~K.

\begin{table}[ht]
\caption{Results for rotation temperature measurements toward the SW absorption.} \label{tab-trot}
\begin{center} \begin{tabular}{lccc}
\hline
Species & Dipole & Date & T$_{\rm rot}$ \\
        & moment & of the     & (K) \\
        & (Debye) & observations & \\
 \hline
C$_2$H         & 0.77 & 2011 &  5.3 $\pm$ 0.1  \\
SO             & 1.54 & 2009 &  5.4 $\pm$ 1.4  \\
HNCO           & 1.58 & 2011 &  9.8 $\pm$ 1.5   \\
HOC$^+$        & 2.77 & 2011 &  5.1 $\pm$ 0.4  \\
H$^{13}$CN     & 2.99 & 2011 &  5.1 $\pm$ 0.2  \\
HC$^{15}$N     & 2.99 & 2011 &  4.1 $\pm$ 0.4  \\
HNC            & 3.05 & 2011 &  4.6 $\pm$ 0.2 $^\dagger$ \\
HN$^{13}$C     & 3.05 & 2011 &  4.8 $\pm$ 0.3  \\
SiO            & 3.10 & 2011 &  6.0 $\pm$ 0.2  \\
c-C$_3$H$_2$-o & 3.43 & 2009 &  5.6 $\pm$ 0.4  \\
c-C$_3$H$_2$-p & 3.43 & 2009 &  5.4 $\pm$ 1.0  \\
HC$_3$N        & 3.73 & 2009 &  6.3 $\pm$ 1.3  \\
H$^{13}$CO$^+$ & 3.90 & 2011 &  5.3 $\pm$ 0.1  \\
\hline
\end{tabular}
\tablefoot{$^\dagger$ Might be affected by opacity effects. Uncertainties are given at 1$\sigma$.}
\end{center} \end{table}

\subsection{Effects of local physical conditions} \label{sect-effectsoflocalconditions}

In order to test the impact of the local physical conditions on the line excitation,
we have run models using the non-LTE molecular radiative transfer code RADEX (\citealt{vdtak07}),
varying parameters such as the kinetic temperature (\Tkin), the H$_2$ density (\nH2),
and the local radiation field.

RADEX uses the escape probability method for a homogeneous medium and is based on given
radiative and collisional transition rates, which are only available for a few species
in a limited range of temperatures and column densities. In this work, we use the
standard files available from the Leiden Atomic and Molecular Database
\footnote{http://home.strw.leidenuniv.nl/$\sim$moldata/} (LAMDA , \citealt{sch05}),
except for HC$_3$N, for which rates extended to 100~K were kindly provided to us by A. Faure
(\citealt{wer07a,wer07b}, and Faure \& Wiesenfeld in preparation).
We consider that H$_2$ molecules are the main collision partners.

\subsubsection{H$_2$ density and kinetic temperature}

We have run RADEX simulations varying \nH2 between 500 and 10$^4$~cm$^{-3}$ for two representative
values of the kinetic temperature \Tkin=50~K and 100~K. These ranges of values encompass previous
estimates of the density ($\sim$2000~cm$^{-3}$, \citealt{hen09}), and of the kinetic temperature
($\sim$80~K, \citealt{hen08}) toward the SW component of PKS\,1830$-$211. Such a value of the kinetic
temperature is typical of \Tkin\ observed in Galactic diffuse clouds ({\em e.g.} \citealt{sno06}).
At this stage and for sake of simplicity, we have set the column densities of the various species
to the values that can be derived directly from Fig.\ref{fig-trot} from the intersection of the
\NLTE\ {\em vs} \Trot\ curves.

\begin{figure*}[h] \begin{center}\includegraphics[width=\textwidth,]{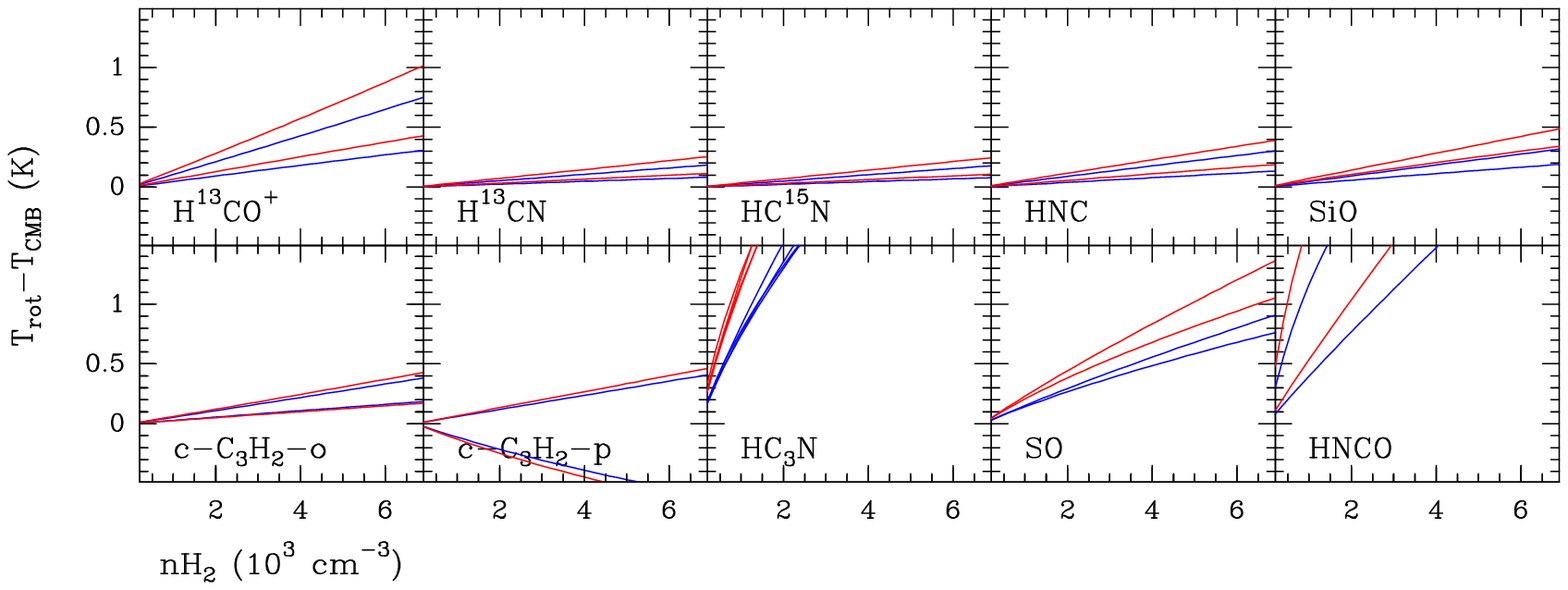}
\caption{RADEX predictions of the rotation temperatures as a function of H$_2$ density for the
transitions of H$^{13}$CO$^+$, H$^{13}$CN, HC$^{15}$N, HNC, SiO, and HNCO listed in Table~\ref{tab-lines}, 
and transitions of c-C$_3$H$_2$, l-C$_3$H$_2$, SO, and HC$_3$N listed in Table~A.1 by \citealt{mul11}.
We assume \Tcmb=5.14~K, \Tkin=50~K
({\em blue}), and \Tkin=100~K ({\em red}). The rotation temperature derived from the observations
of two (or more) transitions is the average of the corresponding rotation temperatures curves.}
\label{fig-radex-nh2}
\end{center} \end{figure*}

The results of our RADEX simulations can be seen in Fig.\ref{fig-radex-nh2}. The curves plotted on
this figure indicate the evolution of the rotation temperature for a given transition as a function
of the H$_2$ density, for HCO$^+$, HCN, HNC, some of their isotopologues, HC$_3$N, SO, HNCO, SiO,
and c-C$_3$H$_2$ (ortho and para forms). The rotation temperature measured from the observation of
two transitions is then the average of the corresponding curves for these two transitions. 
The rotation temperature increases with H$_2$ density, as can be expected when collisional excitation
progressively adds to the radiative excitation from the CMB. This is the case of all transitions, except
for the c-C$_3$H$_2$(para) 3$_{31}$-3$_{22}$, which shows a decrease of \Trot\ with \nH2.

Among the various species investigated here, the most sensitive to collisional excitation are HC$_3$N and
HNCO, for which $\Delta T$=\Trot$-$\Tcmb\ reaches values $>$1~K for \nH2$>$2000~cm$^{-3}$.
On the other hand, the transitions of HCN, HNC and c-C$_3$H$_2$(ortho) (and their isotopic variants
when applicable) have $\Delta T$ of less than a few tenths of a Kelvin for \nH2$\sim$2000~cm$^{-3}$,
and can therefore be considered as good \Tcmb\ indicators.

\subsubsection{Local radiation field}

We now add the excitation from a local radiation field. We consider first the mean
background radiation near the Sun's location in the Galaxy, which consists mostly in dust thermal emission
(we adopt a single-temperature component with $T_{dust}$=23~K, \citealt{wri91}). Non-thermal radiation is
negligible at mm/submm wavelengths. We did not find measurable differences as compared to the previous case
with CMB excitation only. Gradually increasing the dust temperature, we start to see some effects on the
rotation temperature of HNCO for $T_{dust}$$\sim$100~K, while all other molecules do not show any changes.
We thus conclude that the effect of the local radiation on the excitation of the observed lines is
negligible, unless very peculiar conditions prevail.

\subsection{A global solution for the excitation conditions with Monte-Carlo Markov Chains}

In the following, we adopt a global method for solving simultaneously the physical conditions
and \Tcmb. We first describe our working assumptions and the RADEX modeling, and then the
Monte-Carlo Markov Chain approach used to adjust the model to the spectra.

\subsubsection{RADEX model} \label{sec:RADEXmodel}

For a given molecular species, we use RADEX to compute the velocity integrated intensity of
each line present in our spectra, given the background radiation field (\Tcmb), the kinetic
temperature (\Tkin), the number density of H$_2$ molecules (\nH2), the column density of
each molecule (\Ncol) and the width of the molecular line (\DV), which is assumed to be Gaussian.

The physical conditions are possibly changing with time (whenever the barycenter of the continuum
emission drifts). We therefore consider independent physical condition sets for both the 2009
and 2011 data. For simplicity, we assume that the absorbing gas is {\em homogeneous}
and can be characterized, for each epochs, by a single specific kinetic temperature \Tkin\ and
\nH2 density, {\em common} for all molecular species. For the SW line of sight, we thus have the
set of parameters \Tkin/\nH2(2009) and \Tkin/\nH2(2011). The background radiation field is assumed
to be constant and equal to the CMB.

Given the velocity centroid (\V0) and linewidth (\DV), specific for each molecular species, we are
then able to construct a synthetic spectrum, using Eq.~\ref{eq:tau}, covering the observed frequencies.
For each species, we thus have three free parameters (\Ncol, \DV, and \V0) and three tied parameters
(\Tkin, \nH2, and \Tcmb) to fit. In the fit, we include all species with molecular data available
in the LAMDA database, and with at least two, non saturated, lines in the observed spectra.
We end up using: H$^{13}$CO$^+$, H$^{13}$CN, HC$^{15}$N,
c-C$_3$H$_2$ (both ortho and para forms), SiO, CH$_3$CN, SO, HC$_3$N, and HNCO.

\subsubsection{Adjusting the model}

Such a high number of variables is very difficult to constrain using classical techniques.
Our approach here is to adopt a MCMC method, widely used to estimate cosmological parameters
({\em e.g.} \citealt{dun05} and references therein). We use a subclass of the MCMC
methods, the Metropolis-Hastings algorithm to probe the probability distribution of our
problem and study the degeneracies between the different parameters. 

For each parameter, we choose the proposed density, or jumping distribution,
as Gaussian distributions, with variances set initially to the result
of the estimate made in Muller et al. (2011). As the collisional rates are
only tabulated over a particular temperature and density range, we rejected
all jumps going outside the allowed range. The likelihood is simply computed
by $p \propto \exp{ ( - {\chi^2}/{2})}$, with
\begin{equation}
  \label{eq:chi2}
  \chi^2 = \sum_ {\nu=1}^{N} \frac { (s_\nu - s_{\nu,mod})^2
  }{{\sigma_\nu}^2+{\sigma_{\nu,mod}}^2}, 
\end{equation}
where $s_\nu$ is the observed spectrum with its uncertainties $\sigma_\nu$ at
frequency channel $\nu$, and $s_{\nu,mod}$ the modeled spectra with its
uncertainty $\sigma_{\nu,mod}$ in order to take into account the
uncertainty on the continuum illumination $\Delta I_{bg}$ (see Eq.\ref{eq:dIbg}).

After a few initial chains, the variances were set according to the results
of the chains themselves, in order to speed up the convergence of the
adjustment. Note that the final result does not depend on the proposed
density distribution, as long as it is symmetric and the chains have
converged. We used the spectral analysis described in \cite{dun05} to
ensure the convergence of our chains before doing the analysis.

\subsubsection{Result of the Monte-Carlo Markov Chain method}

\begin{figure*} \centering
\includegraphics[width=\textwidth]{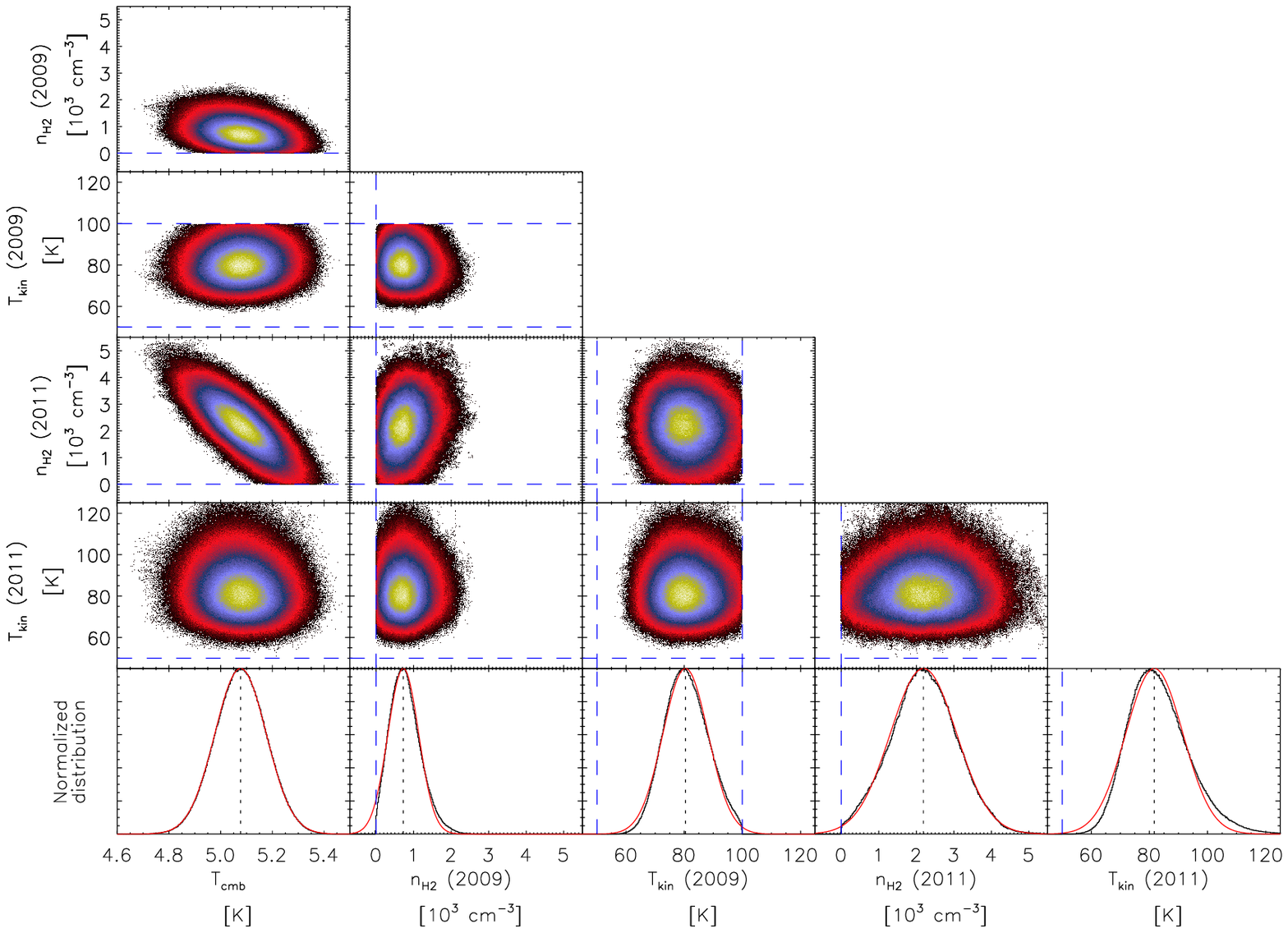}
\caption{One and two dimensional likelihood distributions for the Metropolis-Hasting estimates
of the non-LTE condition toward the SW component of the 2009 and 2011 campaign.
Only a subset of the parameters is represented. The blue dashed lines correspond to the 
common range of temperatures for which collisional rates are available in the LAMDA database
for the considered set of species.
The normalized 1D likelihood distributions
({\em last row}) are fitted by a Gaussian distribution (red continuous line).}
\label{fig:markov_result}
\end{figure*}

One major advantage of the MCMC approach is that it allows us to take all uncertainties
into account when deriving the posterior distribution of the parameters. Most of the
parameters display a posterior Gaussian distribution with median value and 68
percentile error reported in Table~\ref{tab:markov_result}. Fig.~\ref{fig:markov_result}
presents the results for a subset of the parameters, \Tcmb\ and, for the two observing campaigns,
\Tkin\ and \nH2. As expected, there is a slight degeneracy between \Tcmb\ and \nH2, for
both campaigns, where a higher background radiation field could be compensated by a lower
H$_2$ density. However, the number of lines used here allows us to remove partially this
degeneracy, and to put firm constraints on \Tcmb. A percentile analysis (68\% confidence
level) of the final chains, including all lines from all species for both campaigns,
leads to a value of the CMB temperature of
\begin{equation}
  \label{eq:Tcmb}
\Tcmb(z=0.89) =  5.0791^{+0.0993}_{-0.0994}\ {\rm K}.
\end{equation}

\begin{table*} 
\centering
\caption{Results (median and 68 percentile error) of the Metropolis-Hasting estimates of the
physical conditions toward the SW component of the 2009 and 2011 campaigns.}
\label{tab:markov_result}
\begin{tabular}{l c c c c c c}
\hline
Species & \Tcmb  & \nH2            & \Tkin & $\log$ \Ncol & \DV & \V0 \\
       & [K]    & [cm$^{-3}$] & [K]   & [cm$^{-2}$] & [km\,s$^{-1}$] & [km\,s$^{-1}$] \\
\hline
       & $5.08\ ^{+0.10}_{-0.10}$ & & & & & \\
\hline
\textit{(2009)}          &-- & $800\ ^{+400}_{-400}$ & $81\ ^{+ 8}_{- 7}$ & & & \\
CH$_3$CN     & --  & --  &--  & $13.200\ ^{+ 0.016}_{- 0.017}$  & $20.0\ ^{+ 0.9}_{- 0.9}$  & $-2.3\ ^{+ 0.4}_{- 0.4}$ \\
SO           & --  & --  &--  & $13.458\ ^{+ 0.023}_{- 0.023}$  & $19.8\ ^{+ 0.9}_{- 0.9}$  & $-1.8\ ^{+ 0.4}_{- 0.4}$ \\
c-C$_3$H$_2$(o) & --  & --  &--  & $13.711\ ^{+ 0.016}_{- 0.016}$  & $24.1\ ^{+ 0.6}_{- 0.6}$  & $-3.1\ ^{+ 0.3}_{- 0.3}$ \\
c-C$_3$H$_2$(p) & --  & --  &--  & $13.223\ ^{+ 0.019}_{- 0.019}$  & $21.0\ ^{+ 0.7}_{- 0.7}$  & $-1.8\ ^{+ 0.3}_{- 0.3}$ \\
HC$_3$N      & --  & --  &--  & $13.037\ ^{+ 0.027}_{- 0.025}$  & $22.3\ ^{+ 1.2}_{- 1.1}$  & $-5.6\ ^{+ 0.5}_{- 0.5}$ \\
\hline
\textit{(2011)}    & --  & $2200\ ^{+900}_{-900}$ & $82\ ^{+11}_{- 9}$ & & & \\
CH$_3$CN      & --  & --  & --  & $13.530\ ^{+ 0.024}_{- 0.024}$  & $14.6\ ^{+ 1.1}_{- 1.0}$  & $-3.9\ ^{+ 0.5}_{- 0.5}$ \\
H$^{13}$CO$^+$ & --  & --  & --  & $13.264\ ^{+ 0.008}_{- 0.008}$  & $16.2\ ^{+ 0.2}_{- 0.2}$  & $-2.4\ ^{+ 0.1}_{- 0.1}$ \\
H$^{13}$CN     & --  & --  & --  & $13.385\ ^{+ 0.010}_{- 0.010}$  & $17.0\ ^{+ 0.2}_{- 0.2}$  & $-2.8\ ^{+ 0.1}_{- 0.1}$ \\
HC$^{15}$N     & --  & --  & --  & $12.607\ ^{+ 0.023}_{- 0.024}$  & $14.2\ ^{+ 0.9}_{- 0.8}$  & $-3.2\ ^{+ 0.4}_{- 0.4}$ \\
HNCO          & --  & --  & --  & $13.662\ ^{+ 0.029}_{- 0.030}$  & $14.5\ ^{+ 1.2}_{- 1.2}$  & $-4.6\ ^{+ 0.4}_{- 0.4}$ \\
SiO           & --  & --  & --  & $13.369\ ^{+ 0.009}_{- 0.009}$  & $14.6\ ^{+ 0.3}_{- 0.3}$  & $-3.0\ ^{+ 0.1}_{- 0.1}$ \\
\hline
\end{tabular} 
\tablefoot{\Ncol, \DV, and \V0 are the column densities, linewidths (FWHM), and velocity centroids, respectively.}

\end{table*}

\section{Discussion}

\subsection{Cosmological constraints}

Our measurement of \Tcmb\ at $z$=0.89 is interesting because 
{\em i)} it is the most precise measurement at $z$$>$0 to date, reaching a remarkable precision of 2\%;
{\em ii)} the observations of various molecular species allow us to constrain the physical conditions of the absorbing gas,
and limit the possible bias of using only one species;
and {\em iii)} it is derived at a redshift
intermediate between S-Z measurements in galaxy clusters ($z$$<$0.6 so far) and optical/UV quasar absorption
systems ($z$$>$1.5 due to atmospheric cut-off), and with an alternative method (molecular excitation).

\begin{figure}[h] \begin{center}
\includegraphics[width=8.5cm]{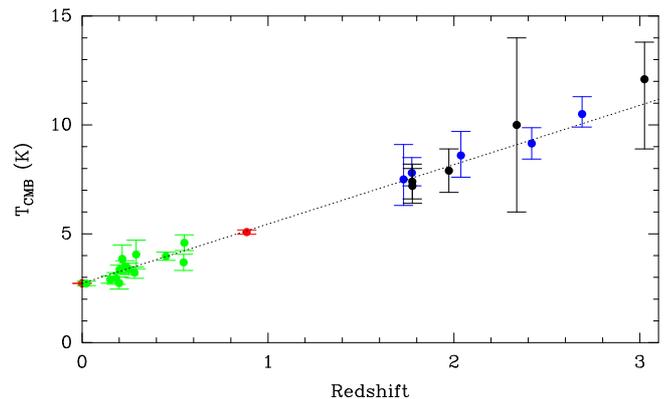}
\caption{Measurements of the CMB temperature as a function of redshift.
Data points in green correspond to S-Z measurements toward galaxy clusters, in black to \ion{C}{i}
and \ion{C}{ii} absorption studies, in blue to CO absorption (see \citealt{not11} and references therein),
and the value derived toward the PKS\,1830$-$211 SW absorption is marked in red. The dotted line
corresponds to the law \Tcmb=$T_0$$\times$(1+$z$).}
\label{fig-Tcmb-vs-z}
\end{center} 
\end{figure}

Including our value of the CMB temperature at $z$=0.89 with other results published in the literature
(see \citealt{not11} and references therein for points plotted in Fig.\ref{fig-Tcmb-vs-z}), we obtain
a tighter constraint on the \Tcmb($z$)=$T_0$$\times$(1+$z$)$^{(1-\alpha)}$ law, $\alpha$=+0.009$\pm$0.019,
compared to the constraint $\alpha$=$-$0.007$\pm$0.027 reported by \cite{not11}.
As far as we can tell, the CMB temperature evolution is consistent with the standard cosmology model, with adiabatic
expansion and $\alpha$=0.

\subsection{Physical conditions of the absorbing gas toward the SW component}

Besides \Tcmb, our MCMC approach also allows us to constrain the physical conditions toward the
SW component, independently for the 2009 and 2011 observations (Table~\ref{tab:markov_result}).
We find that the kinetic temperature is well constrained, mostly owing to the observations of
different K-transitions of CH$_3$CN. Our result of \Tkin$\sim$80~K is comparable for both epochs,
and also similar to the value derived by \cite{hen08} from observations of NH$_3$ inversion
transitions in 2003-2004. Despite probable changes in the line of sight, it seems that the kinetic
temperature of the absorbing gas remains unchanged. The situation is not so clear for the H$_2$
volume density, which is not as well constrained as \Tkin. From 2001-2002 data, \cite{hen09}
estimate \nH2$\sim$2000~cm$^{-3}$, comparable to our 2011 result. By comparison, \nH2 for our 2009
data is roughly half that value, a result that is also found for the integrated opacities
(or column density, all other parameters being equal) between 2011 and 2009. It is possible that
the 2011 line of sight has intercepted a denser, more compact, absorbing component than in 2009.
In any case, based on the observed rotation temperatures and the value of \Tkin, it is clear that
the excitation of the absorbing gas is sub-thermal, and that the density should be moderate
($<$10$^4$~cm$^{-3}$). This is consistent with the absorbing gas being composed of diffuse/translucent
clouds (see \citealt{mul11}).

We note that the increase of absorption intensities in 2011 allows us to detect several new species
toward this galaxy (Table~\ref{tab-lines}): NH$_2$CHO (2 transitions), $^{30}$SiO (J=2-1), HCS$^+$
(whose J=2-1 transition shows up better out of the wing of the c-C$_3$H$_2$ 2$_{12}$-1$_{01}$ line
than in the \citealt{mul11} data), and tentatively HOCO$^+$ (detected at just 4$\sigma$ at rest
frequency 85.531512~GHz).

\subsection{Limits and Perspectives}

Our excitation analysis relies on the simplifying assumption of an homogeneous medium,
characterized by single values of the H$_2$ density and kinetic temperature common for all
molecules, when it is an observational fact that Galactic clouds are highly structured and
turbulent. \cite{hen08} interpret NH$_3$ observations to show two components of absorbing
gas, one at approximately 80~K occupying 80-90\% of the column, and a second at kinetic
temperature greater than 100~K occupying 10-20\% of the column. The warmer component would
be even more diffuse and more subthermally excited, if in pressure equilibrium with the
denser and cooler phase. The typical scale length corresponding to a H$_2$ column density
of 2$\times$10$^{22}$~cm$^{-2}$ (see \S4.2 by \citealt{mul11} for the SW line of sight) and
a density \nH2=2000~cm$^{-3}$ is $\sim$3~pc. Considering VLBI measurements of the size of
the continuum images ($\sim$2pc in the image plane at $z$=0.89, \citealt{jin03}), the high
(near 100\%) covering factor of the absorbing clouds toward the SW line of sight, and the
time variation of the absorption profiles, we conclude that the extent of the clouds should
be of the order of a few parsecs. Hence, the bulk of the absorption should be dominated by
the cool component with \Tkin $\sim$80~K. We thus estimate that the contribution from the
warm component does not affect significantly our measurement of \Tcmb, and our hypothesis
of an homogeneous medium is reasonable.

Although the derived value of the background radiation temperature is robust in our analysis,
some neutral species like SiO and HNCO show slightly elevated rotation temperatures, suggesting
that they may require higher densities than the ions. This could be explained most easily if
there is chemical stratification in the absorbing column. On the other hand, some of the crucial
collision rates are poorly known, especially when the ortho form of H$_2$ is the main collision partner. 

We also assume that the continuum source is fully covered ($f_c$=1 in Eq.\ref{eq:tau}). Nevertheless,
this is not a severe assumption for optically thin lines, and recent ALMA observations of the
ground-state line of ortho water, resolving both NE and SW images of the quasar, show that the SW
continuum source covering factor is indeed close to unity (Muller et al., in prep.), at least at the time
of the observations in 2012.

Finally, we have considered H$_2$ molecules only as collision partners. As shown by \cite{lis12} for the
Galactic interstellar medium, molecular ions with large dipole moments like HCO$^+$ are most sensitive to
collisional excitation by electrons. Electron-impact begins to compete with neutral collisions when the
fractional ionization $n(e)/n_{\rm H} \gtrsim 10^{-4}$. The observed rotational temperatures of H$^{13}$CO$^+$
and HOC$^{+}$ are both close to the expected \Tcmb=5.14~K, suggesting that the contribution from collisions
with electrons is not significant.

The robustness of the \Tcmb\ measurement toward PKS\,1830$-$211 could be improved by investigating the
different lines of sight to the lensed images of the background quasar. New observations when the NE
absorption becomes stronger again ({\em e.g.} as in the period 1999--2003, see \citealt{mul08}), with
angular resolution sufficient to resolve the NE and SW images (separated by 1$\arcsec$), would provide
an interesting independent \Tcmb\ measurement.

Additional transitions/molecules could be observed (within a short time interval to minimize time
variations) to place tighter constraints on the physical conditions. However, and very importantly,
collisional rates (with H$_2$) are still lacking for a large number of molecules. This work is a
good illustration of the importance and needs  in astrophysical studies of molecular spectroscopy
and collisional data (see {\em e.g.} \citealt{vdtak11} for a short review on the field).
	
We have shown that using molecular transitions as diagnostics of physical conditions toward molecular
absorbers is a powerful method to measure \Tcmb. Unfortunately, there is only a small number of $z$$>$0
millimetre-wave band molecular absorbers known to date (see a review by \citealt{com08}, and also
\citealt{cur11}). Due to the brightness of the background continuum, the large column density of absorbing
gas, the large number of detected molecules, and the multiple independent lines of sight, the absorber
toward PKS\,1830$-$211 is a prime target for a precise measurement of \Tcmb. Another relatively well-suited
target could be the $z$=0.68 absorber located toward the quasar B\,0218+357 (\citealt{wik95}). The physical
conditions have already been explored using transitions of ammonia (\citealt{hen05}) and formaldehyde
(\citealt{jet07, zei10}), but a thorough investigation combining several molecules remains to be done for
this source, in order to derive a precise measurement of \Tcmb\ at $z$=0.68. A potentially severe difficulty
is the small angular separation of $\sim$0.3$\arcsec$ between the two gravitationally lensed images of the quasar. 

In the (near) future, new $z$$>$0 molecular absorbers might be discovered ({\em e.g.} with the Atacama Large
Millimeter-submillimeter Array or the Karl G. Jansky Very Large Array), providing more chances to test the
\Tcmb\ evolution with redshift using molecular excitation analysis.

\section{Summary and conclusions} \label{Conclusion}

We have performed a multi-transition excitation analysis of various molecular absorption lines toward
the radio-mm molecular absorber PKS\,1830$-$211, with the aim to precisely determine the cosmic microwave
background temperature at $z$=0.89. We use new 7 and 3~mm ATCA observations obtained in 2011, complemented
with 2009 data from our ATCA 7~mm survey (\citealt{mul11}). Our results can be summarized as follows:

\begin{itemize}

\item The rotation temperatures of the observed molecules are all close to $\sim$5~K,
comparable to the value \Tcmb=5.14~K predicted from the adiabatic expansion of the Universe.
This suggests that the molecular excitation is mostly radiatively coupled with the CMB,
and that an accurate value of \Tcmb\ can be derived.

\item We combine a Monte-Carlo Markov Chain approach with non-LTE radiative transfer predictions
from the RADEX code to solve for the physical conditions of the absorbing gas, taking into account
time variations of the quasar ({\em i.e.} possible changes of the sightline) between
the 2009 and 2011 observations.
For each epoch, we assume that the absorbing screen is homogeneous, and characterized
by a single kinetic temperature and H$_2$ density.
Under this (reasonable) assumption, and with the molecular data available,
we determine a CMB temperature of 5.08$\pm$0.10~K (68\% confidence level)
at $z$=0.89, which is the most precise measurement of \Tcmb\ at $z$$>$0 to date.

\item Combining this result with other measurements published in the literature, we refine
the evolution of the CMB temperature as a function of redshift,
$T$=$T_0$$\times$(1+$z$)$^{(1-\alpha)}$, finding $\alpha$=0.009$\pm$0.019, {\em i.e.} a quasi-linear
dependence to the redshift. As far as we can tell, this is consistent with adiabatic expansion of the Universe.

\item The absorbing gas has properties comparable to Galactic diffuse/translucent clouds, with a 
kinetic temperature of $\sim$80~K and a H$_2$ density of the order of 10$^3$~cm$^{-3}$.

\item We report the detection of several new species toward PKS\,1830$-$211:
NH$_2$CHO, $^{30}$SiO, HCS$^+$ and tentatively HOCO$^+$. The weak lines of these species could
be detected in 2011 data toward the SW component owing to the general increase in absorption
depth of all species by a factor of two compared to our previous observations in 2009.

\end{itemize}

This work emphasizes the extreme importance of accurate molecular collisional data
for astrophysical studies.

\begin{acknowledgement}

We thank the anonymous referee and the Editor, Malcolm Walmsley, for useful comments and suggestions
to improve the presentation of the paper.
This work makes intense use of the LAMDA Database and RADEX program, we thank all the people
who contributed to these projects. We thank A. Faure for providing us with an extended molecular
data file for HC$_3$N. AB would like to thank M. Douspis for fruitful discussions and endless candies.
The Centre for All-Sky Astrophysics is an Australian Research Council Centre of Excellence,
funded by grant CE110001020.

The Australia Telescope Compact Array is part of the Australia Telescope which is funded by the
Commonwealth of Australia for operation as a National Facility managed by CSIRO.

Based on observations carried out with the IRAM Plateau de Bure Interferometer. IRAM is supported by
INSU/CNRS (France), MPG (Germany) and IGN (Spain).

\end{acknowledgement}

\begin{appendix}

\section{Spectra and fit results}

\begin{figure*}[h] \begin{center}
\includegraphics[width=\textwidth]{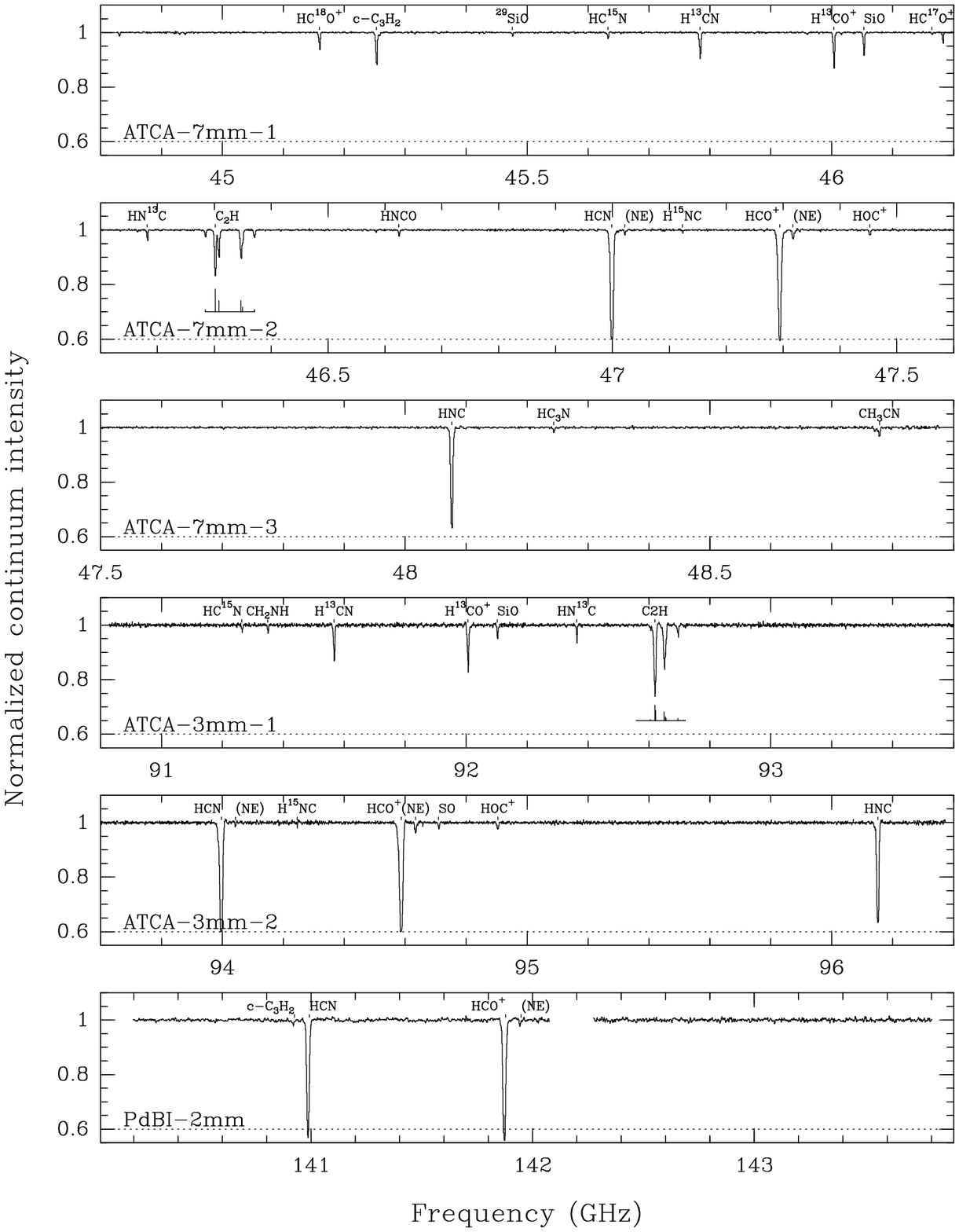}
\caption{Spectra from ATCA and PdBI observations. The frequency scale is in the observer frame. The spectra are normalized
to the total flux density corresponding to the sum of the NE and SW images.}
\label{fig-spec-full}
\end{center} \end{figure*}

\begin{figure*}[h] \begin{center}
\includegraphics[width=\textwidth]{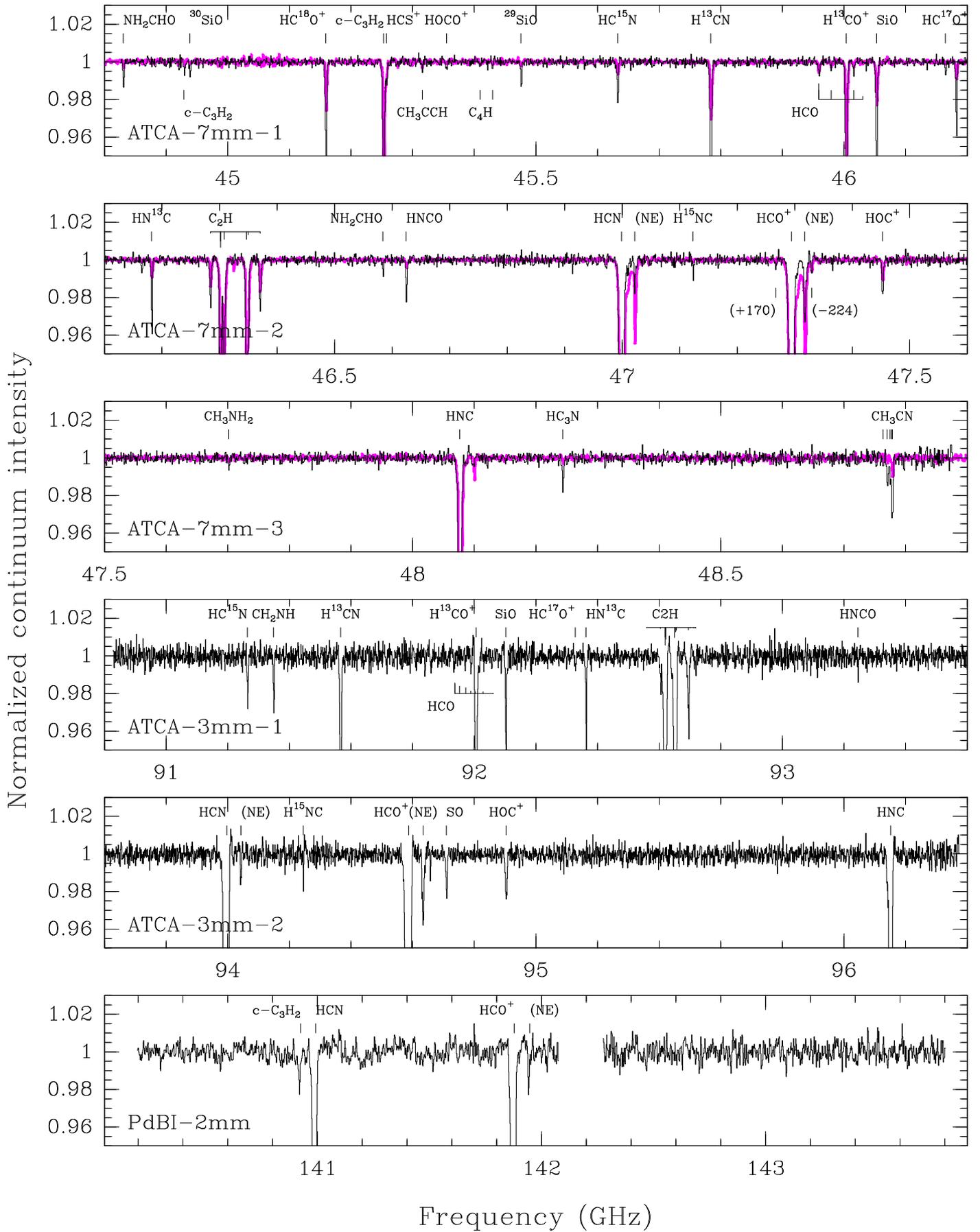}
\caption{Same as Fig.\ref{fig-spec-full}, but zoomed in for weak lines and with 2009/2010 averaged 7~mm spectrum
overlaid (magenta).}
\label{fig-spec-zoom}
\end{center} \end{figure*}

\clearpage
\onecolumn
\begin{longtable}{rlrrrrr}
\caption{Parameters and results of the global fit of the spectra for non-saturated lines detected toward the SW component in 2011 ATCA observations.} \label{tab-lines} \\
\hline
 
Rest Freq. & Transition  & \multicolumn{1}{c}{$S_{ul}$ $^a$} & \multicolumn{1}{c}{$E_l/k_B$ $^b$} & \multicolumn{1}{c}{V$_0$ $^c$} & \multicolumn{1}{c}{$\Delta V$ $^d$} & \multicolumn{1}{c}{$\int \tau $d$V$ $^e$} \\
\multicolumn{1}{c}{(MHz)}      &             &          &  \multicolumn{1}{c}{(K)} & (km~s$^{-1}$) & (km~s$^{-1}$) & \multicolumn{1}{c}{(10$^{-3}$ km~s$^{-1}$)}\\
 
\hline
\endfirsthead
\caption{Continued.} \\
\hline
Rest Freq. & Transition  & \multicolumn{1}{c}{$S_{ul}$} & \multicolumn{1}{c}{$E_l/k_B$} & V$_0$ & $\Delta V$ & $\int \tau $d$V$\\
\multicolumn{1}{c}{(MHz)}      &             &          &  \multicolumn{1}{c}{(K)} & (km~s$^{-1}$) & (km~s$^{-1}$) & (10$^{-3}$  km~s$^{-1}$)\\
 
\hline
\endhead
 
\endfoot
 87316.898&              C$_2$H N=1-0 & 1.00 &  0.0 &  $-$0.5 (0.1)& 18.9 (0.1) & 28234 (280) \\
174663.199&              C$_2$H N=2-1 & 2.00 &  4.2 & -- $^t$           & -- $^t$          & 37574 (486) \\
 86339.922&          H$^{13}$CN J=1-0 & 1.00 &  0.0 &  $-$2.7 (0.1) & 16.8 (0.2) & 4966 (69) \\
172677.851&          H$^{13}$CN J=2-1 & 2.00 &  4.1 & -- $^t$ & -- $^t$ & 6393(141)\\
 86054.966&          HC$^{15}$N J=1-0 & 1.00 &  0.0 & -- $^t$ & -- $^t$ &  945(42)\\
172107.957&          HC$^{15}$N J=2-1 & 2.00 &  4.1 & -- $^t$ &-- $^t$&  943(99)\\
 90663.568&                 HNC J=1-0& 1.00&  0.0&   0.17 (0.03) & 19.8 (0.1) & 43339 (1130) \\
181324.758&                 HNC J=2-1& 2.00&  4.4 &--  $^t$ &--  $^t$ & 45856 (1347) \\

 87090.825&          HN$^{13}$C J=1-0& 1.00&  0.0&-- $^t$&-- $^t$& 1515 (50) \\

174179.411&          HN$^{13}$C J=2-1& 2.00&  4.2&-- $^t$&-- $^t$& 1778 (86) \\
 88865.715&          H$^{15}$NC J=1-0& 1.00&  0.0&-- $^t$&-- $^t$&  300 (50) \\
 86754.288&      H$^{13}$CO$^+$ J=1-0& 1.00&  0.0&  $-$2.7 (0.1) & 15.6 (0.1) & 6089 (79) \\
173506.700&      H$^{13}$CO$^+$ J=2-1& 2.00&  4.2&-- $^t$&-- $^t$& 8072 (153) \\
 85162.223&      HC$^{18}$O$^+$ J=1-0& 1.00&  0.0&-- $^t$&-- $^t$& 2581 (46) \\
 87057.535&      HC$^{17}$O$^+$ J=1-0& 1.00&  0.0&-- $^t$&-- $^t$&  262 (40) \\
174113.169&      HC$^{17}$O$^+$ J=2-1& 2.00&  4.2&-- $^t$&-- $^t$&  231 (70) \\
 89487.414&             HOC$^+$ J=1-0& 1.00&  0.0&  $-$0.2 (0.5) & 19.3 (1.1) & 1057 (62) \\
178972.051&             HOC$^+$ J=2-1& 2.00&  4.3 & -- $^t$ & -- $^t$ & 1283 (92) \\
 86670.760&                 HCO N$_K$=1$_{01}$-0$_{00}$ J=3/2-1/2& 1.00&  0.0&  $-$1.7 (1.0)&18.8 (2.4)&  833 (62) \\
172266.853&            CH$_2$NH 2$_{11}$-2$_{02}$& 7.40&  9.2 &  $-$4.1 (0.6) &  9.8 (1.4) &  812 (61) \\
 89956.068&     CH$_3$NH$_2$(p) 1$_{10}$-1$_{01}$& 6.00&  2.1 &  $-$4.8 (1.5) & 12.3 (3.2) &  223 (53) \\
 85338.893&      c-C$_3$H$_2$-o 2$_{12}$-1$_{01}$& 4.50&  2.3 &  $-$1.9 (0.1) & 17.8 (0.2) & 6973 (85) \\
 84727.696&     c-C$_3$H$_2$(p) 3$_{22}$-3$_{13}$ & 0.96& 12.1&-- $^t$&-- $^t$&  244 (38) \\
 91958.726&            CH$_3$CN J$_K$=5$_4$-4$_4$& 3.60&123.1 &$-$3.9 (0.5) & 14.6 (1.1)&-- \\
 91971.130&            CH$_3$CN J$_K$=5$_3$-4$_3$&12.80& 73.1 &-- $^t$&-- $^t$&-- \\
 91979.994&            CH$_3$CN J$_K$=5$_2$-4$_2$& 8.40& 37.4 &-- $^t$&-- $^t$&-- \\
 91985.314&            CH$_3$CN J$_K$=5$_1$-4$_1$& 9.60& 16.0 &-- $^t$&-- $^t$&-- \\
 91987.088&            CH$_3$CN J$_K$=5$_0$-4$_0$&10.00&  8.8 &-- $^t$&-- $^t$&-- \\
 87925.237&                HNCO 4$_{04}$-3$_{03}$& 4.00&  6.3 &  $-$4.5 (0.4) & 14.2 (1.0) &  845 (51) \\
175843.695&                HNCO 8$_{08}$-7$_{07}$& 8.00& 29.5 &-- $^t$&-- $^t$&  257 (81) \\
 86846.960&                 SiO J=2-1& 2.00&  2.1&  $-$3.0 (0.1) & 14.4 (0.2) & 3612 (60) \\
173688.310&                 SiO J=4-3& 4.00& 12.5 & -- $^t$ & -- $^t$ & 1944 (96) \\
 85759.199&          $^{29}$SiO J=2-1& 2.00&  2.1 & -- $^t$ & -- $^t$ &  541 (37) \\
 84746.170&          $^{30}$SiO J=2-1& 2.00&  2.0 & -- $^t$ & -- $^t$ &  287 (35) \\
 85347.890&             HCS$^+$ J=2-1& 2.00&  2.0&  $-$1.5 (0.0) $^f$ & 15.0 (1.4) &  542 (45) \\
 85531.512&   HOCO$^+$ 4$_{04}$-3$_{03}$ & 4.00 & 6.2 & 0.5 (2.3) &   17.8 (5.5) & 181 (48) \\
 84542.329&           NH$_2$CHO 4$_{04}$-3$_{03}$&12.00&  6.1 &  $-$4.4 (0.8) & 15.0 (0.0)$^f$ &  507 (52) \\
 87848.873&           NH$_2$CHO 4$_{13}$-3$_{12}$&11.23&  9.3 &-- $^t$ &-- $^t$ &  290 (42) \\
178605.403&                  SO J$_K$=5$_4$-4$_3$& 4.91 & 15.9 &  $-$3.4 (0.7) & 13.7 (1.6) &  843 (85) \\
 90979.023&             HC$_3$N J=10-9&10.00& 19.6&  $-$2.5 (0.7) & 14.6 (1.6) &  680 (66) \\
\hline \end{longtable}
\tablefoot{$^a$ Line strength. $^b$ Energy of the lower level of the transition. $^c$ Line centroid. $^d$ FWHM linewidth. $^e$ Integrated opacity.
$^t$ The parameter was tied with the precedent entry. $^f$ The value was fixed.
For some species, the fit results might be slightly different (although within uncertainties) to the
values given in Table~\ref{tab:markov_result}, due to the different fitting procedure with tied parameters.}

\twocolumn

\end{appendix}

\end{document}